# Frontier Safety Policies for AI Emergency Preparedness in China

JAMES ZHANG*, Tsinghua University, China and ERA Fellowship, UK
MILES KODAMA, Williams College, USA
ZONGZE WU, Tsinghua University, China
MICHAEL CHEN, University of Oxford, UK
YUE ZHU, Tongji University, China
GENG HONG, Fudan University, China

Chinese authorities now recognize artificial intelligence incidents as a potential source of catastrophe, comparable to natural disasters, cyberattacks, and epidemics. Current regulations require pre-deployment assessments and incident reporting, but mechanisms for emergency preparedness – proactive measures to reduce catastrophic risk throughout the R&D process – remain underdeveloped. This paper suggests frontier safety policies (FSPs), frameworks where developers pre-commit to safety measures conditional on crossing specific dangerous capability thresholds. We show that the core structure of FSPs maps onto the prevention and monitoring phases of China's four-phase emergency response framework, and recommend that Chinese regulators require catastrophic risk evaluations and mandate that frontier AI developers maintain FSPs. We also compare China's regulatory landscape with the EU General-Purpose AI Code of Practice and with recent U.S. state AI legislation, showing emerging global convergence around the use of FSPs.

CCS Concepts: • **Computing methodologies → Artificial intelligence**; **Machine learning**; • **Social and professional topics → Computing / technology policy**; **Governmental regulations**.

Additional Key Words and Phrases: AI Governance, China, Emergency Response, Catastrophic Risk, Frontier Safety Policy, Dangerous Capability Evaluation, AI Regulation, Risk Management

## 1 Motivating Emergency Response Measures

China's top authorities have recently called for the country to prepare for emergencies caused by AI. In an April 2025 study session of the Politburo, President Xi Jinping called for China to "establish systems for technical monitoring, early risk warning and emergency response" to guarantee AI's "safety, reliability and controllability" [78]. The State Council subsequently published a whitepaper on national security, saying that China aims to establish "agile governance, tiered management, and rapid, effective response" for emerging security-relevant

*JZ, MK, ZW, and MC contributed equally to this work.

Authors' Contact Information: James Zhang, james.zhang@sc.tsinghua.edu.cn, Tsinghua University, Beijing, China and ERA Fellowship, Cambridge, UK; Miles Kodama, mmk4@williams.edu, Williams College, Williamstown, Massachusetts, USA; Zongze Wu, zongze_wu@mail.tsinghua.edu.cn, Tsinghua University, Beijing, China; Michael Chen, michael.chen@linacre.ox.ac.uk, University of Oxford, Oxford, UK; Yue Zhu, yue_zhu@tongji.edu.cn, Tongji University, Shanghai, China; Geng Hong, ghong@fudan.edu.cn, Fudan University, Shanghai, China.

technologies such as AI [71]. In September 2024, an international group of AI experts including Andrew Yao, Zhang Hongjiang, and Xue Lan signed a statement calling upon governments to establish institutions for AI "emergency preparedness" [36]. By December 2024, the National Technical Committee 260 (TC260) had released a draft guide for "Generative Artificial Intelligence Service Security Emergency Response," which was officially updated and issued in September 2025 [52]. Overall, a broad consensus is emerging that China should prepare for potential future emergencies caused by AI.

**China classifies AI security incidents alongside natural and national disasters**. Authorities have made it clear that the AI emergencies they envisage include catastrophes more severe than those ordinarily caused by malfunctioning software. The new National Emergency Response Plan released in February 2025 put "artificial intelligence security" incidents next to earthquakes, network data security, and infectious diseases, on a list of areas requiring "mass monitoring and prevention" [14]. This indicates that Chinese authorities are seriously considering *catastrophic* AI risks, defined as risks of exceptionally destructive events (catastrophes) that persistently affect many people over a wide geographic area [12]. An AI catastrophe would involve loss of life, mass injury, or loss of property on the scale of a major natural disaster or worse.[1]

One widely acknowledged way that AI could cause a catastrophe as devastating as a natural disaster is by giving malicious actors access to dangerous knowledge that enables them to build weapons of mass destruction. Another way is a loss of control event, where an AI evades its creators' control and acquires resources autonomously to replicate itself and amass power. TC260's AI Safety Governance Framework acknowledges both threat models; its 2025 update calls for enhancing consensus on collaboratively addressing AI loss-of-control risks and adds "prevent loss of control" as a governance principle [46, 51]. In this paper, an "emergency" is generally something like either of these two catastrophic outcomes, and emergency preparedness consists of measures to make such outcomes less likely. Analogizing to risks from fire, we are more interested in fire-resistant building codes (that prevent fires from happening) than fire escapes (that help mitigate the losses if a fire takes place).

**China's emergency management system provides a foundation for AI risk governance**. Although catastrophic risks from AI are novel and unprecedented, China's emergency management system features an established and institutionalized architecture. Since the early 2000s, China's "One Plan, Three Systems" has served as a unified framework for emergency response, defining clear responsibilities, enabling rapid resource mobilization, and mandating emergency preparedness [75]. A key component of the framework is the Emergency Response Law of the People's Republic of China, which divides the emergency response process into four phases: prevention and preparedness, monitoring and early warning, response and rescue, and recovery and reconstruction. This four-stage system has been used to respond to natural disasters, severe accidents, and public health incidents; recently, TC260 has adopted such a framework for AI emergency response [52].

While the four-phase loop provides structure for emergency response, much work remains to be done on its technical implementation. In January 2025, TC260 released a draft on the nation's "AI Safety Standards System," outlining current and future technical standards for operationalizing its AI Safety Governance Framework [50]. To complement this work, we examine the model of frontier safety policies widely used by leading AI companies and increasingly by governments worldwide.

We suggest that a system based on dangerous capability evaluations, thresholds, and pre-planned safety mitigations could constitute part of the first two phases of this emergency response loop. AI developers can prepare for emergencies by announcing how they will test their models for dangerous capabilities, what threshold capabilities would trigger heightened response, and what security measures they will impose on models at each threshold. The developers can then monitor for emergencies by performing the prescribed dangerous capability tests on their models at set intervals, providing a real-time indicator of catastrophic risk. Our focus throughout is on frontier generative AI, consistent with the scope of the regulatory instruments analyzed.

[1]By this definition, an AI catastrophe could be expected to cause at least dozens of casualties or billions of dollars worth of damage. This is consistent with the definition of the most severe incident tier in China's National Natural Disaster Relief Emergency Plan and with the definition of "catastrophic risk" in California's Senate Bill 53 [10, 70].

In §2, we discuss existing frontier AI regulations in China and how they could contribute to emergency preparedness. §3 then covers prevailing global practices in AI emergency preparedness, looking both at formal AI regulations in other jurisdictions and at voluntary industry standards. §4 examines implementations for each phase of the emergency response process. In §5, we review related scholarship on frontier safety policies and Chinese AI governance. Finally, §6 concludes by highlighting key considerations for policymakers and suggesting how existing regulations and drafts might be extended.

## 2 Regulating Frontier AI in China

In this section, we examine the landscape of AI regulation and emergency preparedness in China. Namely, we review foundational regulatory instruments and voluntary initiatives carried out by key players in the country's AI ecosystem.

### 2.1 Government Regulatory Frameworks

**Interim Measures for the Management of Generative AI Services** (August 2023). Issued by the Cyberspace Administration of China (CAC), the "Interim Measures" are China's first comprehensive regulation for public-facing generative AI systems and reflect a preemptive and compliance-oriented governance approach [24]. The Interim Measures require all providers of generative AI models with public opinion or social mobilization capabilities to register their models with regulators and pass pre-deployment security assessments (Article 17), ensuring formal reviews before release. To pass, providers must show that their models uphold core socialist values, avoid discrimination, deal fairly in commercial settings, respect individuals' rights, and are adequately transparent and reliable (Article 4). These Measures create a foundation for emergency preparedness by establishing government oversight of frontier AI development and mandating pre-deployment evaluations.

**AI Safety Governance Framework** (draft released September 2024, updated September 2025). Published jointly by TC260 and China's National Computer Network Emergency Response Technical Team (CNCERT), the Framework is a non-binding guidance document that adopts a risk-based regulatory philosophy, setting forth high-level governance strategies and technical measures for three categories of AI risk: inherent, application-based, and societal [46, 51]. The identified risks include the proliferation of weapons of mass destruction (biological, chemical, nuclear, and cyber) and the loss of human control; the Framework also acknowledges the misuse potential stemming from open-source AI models. To inform future AI regulations, it introduces a risk classification and grading system that assesses risks by application scenario, scale, and level of intelligence. The Framework serves as the foundation for subsequent standard-setting efforts, as referenced by TC260's draft on the "AI Safety Standards System" (January 2025) [50].

**Emergency Response Guidelines for Security of Generative AI Services** (draft released December 2024, updated September 2025). The "Practice Guidelines" mark the country's first attempt to outline a structured emergency response playbook targeting generative AI [48, 52]. This guidance classifies security incidents into ten categories (drawn from standard GB/T 20986-2023) and defines four levels of incident severity. Among the covered incidents are AI-generated cyberattacks, model tampering and backdoors, along with many other concerns; the guidance also recommends automated real-time monitoring of model inputs and outputs. It further proposes technical and policy emergency response protocols across the four stages: prevention and preparedness, monitoring and early warning, response and rescue, and recovery and reconstruction. The Guidelines call for generative AI service providers to invest in emergency preparedness before emergencies strike, monitor continuously to identify emergencies as early as possible, act swiftly to address identified emergencies, and assess their own responses after emergencies have passed.

**GB/T 45654-2025: Basic Security Requirements for Generative Artificial Intelligence Services** (April 2025). This recommended standard establishes prescriptive technical specifications for generative AI model providers across the entire model lifecycle, from training data collection to pre-/post-training evaluation to post-deployment monitoring [47]. It addresses 31 risk categories spanning content safety, discrimination, commercial fairness, individual rights, and service reliability. The standard operationalizes the safety goals outlined in the Interim Measures by specifying concrete criteria for pre-deployment evaluations and further requires model providers to implement mechanisms for rapid response and feedback to emerging emergencies.

**Cybersecurity Law** (updated October 2025). This wide-ranging cybersecurity law, enacted in 2016 and updated in October 2025, includes new language on AI risk monitoring, assessment, and security oversight. The update also endorses the use of AI to assist with cybersecurity protection [77]. A November 2025 report from the China Academy of Information and Communications Technology (CAICT), a government-affiliated research institute, provides further analysis of AI security issues, evaluating fifteen code generation models from Zhipu, DeepSeek, and Qwen and finding two to be high risk. One concern was the "success rate for non-professionals in generating complete attack code." The report also refers to existing AI security risk assessment frameworks from the EU, UK, and major AI developers, and highlights challenges in securing AI infrastructure, training data, and model robustness [17].

China's AI safety regulation currently centers on tightening oversight at entry points, segmenting risks early, embedding safeguards across the lifecycle, and formalizing incident management protocols. While these measures primarily target content-level and security-related harms rather than catastrophic risk prevention, they establish important precedents. Requirements for pre-deployment assessment, continuous monitoring, and rapid response to identified risks provide a foundation that could be extended to address more severe, system-level threats from AI. The quantitative evaluation thresholds and detailed technical specifications set out in GB/T 45654-2025 further demonstrate regulators' willingness to impose capability-based AI safety standards.

## 2.2 Self-Governance Initiatives

Beyond government regulations, China's AI ecosystem has begun developing voluntary safety frameworks that could enhance emergency preparedness. In July 2025, the nonprofit Shanghai AI Laboratory, in collaboration with the social enterprise Concordia AI, released a comprehensive Frontier AI Risk Management Framework, proposing systematic approaches to identifying, evaluating, and mitigating catastrophic AI risks [64]. The Framework is meant to serve as an exemplar for general-purpose AI model developers and stakeholders to adopt.

**Frontier AI Risk Management Framework** (July 2025). The joint effort signals a shift in the ecosystem's approach to safety, moving beyond content moderation to addressing catastrophic risks. The Framework establishes a three-tiered risk classification system with "yellow lines" serving as early warning indicators and "red lines" marking unacceptable risk thresholds. When models approach yellow lines, the Framework calls for heightened protective measures and additional oversight before proceeding. Crossing red lines can trigger restrictions on further development or public release of the model. Critically, the Framework specifically addresses the catastrophic risks that authorities have identified as requiring emergency preparedness, including detailed thresholds for cyber offense capabilities, biological threats, and loss of control scenarios.

The Framework's approach to capability evaluation is particularly relevant for emergency preparedness. It recommends continuous risk analysis throughout the AI development lifecycle, including pre-development capability prediction through scaling laws, pre-deployment safety assessments, and post-deployment monitoring. For biological risks, for example, the Framework establishes specific red lines such as "lowering barriers to acquiring and proliferating harmful biological agents" and prescribes corresponding evaluations using benchmarks like WMDP-Bio [42] and LAB-Bench [41]. This creates a direct link between capability measurement and emergency response planning.

**AI Safety and Emergency Preparedness in the Industry.** Meanwhile, many AI companies in China have made voluntary commitments relevant to emergency preparedness. Seventeen leading Chinese AI companies—including Baidu, Alibaba, DeepSeek, and ByteDance—signed the "Artificial Intelligence Safety Commitments" in December 2024 under the auspices of the state-backed China AI Industry Alliance (AIIA) [18]. The signatories pledged to "conduct safety assessments before models and services are made public" and to "perform continuous testing throughout the model's lifecycle, including security assessments, red teaming, and alignment checks." They further commit to establishing "management mechanism[s] for security vulnerabilities and other risks, taking timely measures for remediation," as is required for rapid, effective emergency response.

These Safety Commitments were updated at the World AI Conference in July 2025 under a plenary session convened by the China AI Safety and Development Association [20]. The new Commitments call for signatories to "strengthen the assessment of risks related to the abuse of AI systems in frontier fields, and prevent potential risks of their abuse in high-risk scenarios," in a possible nod to preventing catastrophic misuse. Five Chinese AI companies that were not signatories to the December 2024 Commitments signed the updated Commitments. One company, DeepSeek, has reportedly conducted evaluations of frontier risks of its AI models, including self-replication and cyber offense capabilities, although the evaluations were not publicly disclosed [23].

## 3 International Approaches to AI Emergency Response

As China advances its approach to AI emergency response, other jurisdictions worldwide are independently reaching similar conclusions about the types of catastrophic risks and the safety infrastructure needed to manage them. Recent regulatory developments in the European Union and the United States reflect an emerging consensus on core emergency preparedness mechanisms that include capability-based risk assessment, tiered response protocols, and rapid incident reporting systems. These frameworks reflect principles similar to China's policy direction and could be leveraged to enhance its implementation approach.

### 3.1 The European Union AI Act: General-Purpose AI Code of Practice

The EU's Code of Practice for General-Purpose AI Models (July 2025) implements many of the same emergency preparedness goals that Chinese frameworks identify as priorities [27]. The Code is voluntary, but under Article 55(2) of the AI Act, providers of general-purpose AI models with systemic risk may use it to demonstrate compliance with the Act's binding obligations, while non-signatories must demonstrate equivalent compliance through alternative means [55]. That systemic risk designation is part of the Act's framework on model size and capability, which we focus on here, as opposed to its separate high-risk system framework on deployment contexts. The Code recognizes four catastrophic risk categories requiring special attention: risks from chemical, biological, radiological, and nuclear (CBRN) attacks; loss of control scenarios where humans cannot reliably direct or shut down AI systems; sophisticated cyber-attacks; and harmful manipulation of human behavior.

To prepare for these four types of emergencies, the Code commits general-purpose AI service providers to establish comprehensive "Safety and Security Frameworks" before placing models on the market. These frameworks should define "systemic risk tiers" based on measurable model capabilities, with at least one tier representing capabilities not yet achieved by current models. This forward-looking approach enables proactive emergency preparedness, requiring providers to estimate when they might develop models surpassing the highest risk tier so authorities can plan response measures in advance. The recommendation to define "trigger points" for additional evaluations throughout the model lifecycle creates natural checkpoints where emergency preparedness can be assessed and updated.

Particularly relevant for emergency response is the Code's incident reporting framework, which details concrete timelines based on severity: two days for incidents causing "serious and irreversible disruption of critical infrastructure," five days for serious cybersecurity breaches, and graduated timelines for other harms. This tiered approach, combined with the principle that reporting a serious incident is not an admission of wrongdoing, encourages rapid information sharing that could prove crucial during an AI emergency. The Code also establishes continuous monitoring through "lighter-touch model evaluations at appropriate trigger points" defined by metrics such as training compute, development stages, and user access.

### 3.2 State-level Emergency Preparedness in the United States

While the U.S. does not yet have comprehensive federal AI regulation, individual states have begun establishing emergency response frameworks. Namely, both California and New York—two of the largest and most innovative U.S. states—have passed transparency-centered legislation that requires major AI developers to disclose their approach to frontier safety and report incidents to the government. These state AI bills generally cannot be preempted without an act of Congress, notwithstanding a December 2025 executive order [15, 28].

**California Senate Bill 53 (SB53)**. Signed into law in September 2025, SB53 requires every large frontier AI developer to publish and follow a framework detailing how they manage any catastrophic risks posed by their models [10]. These frameworks must describe how the developer assesses their models' dangerous capabilities and what precautions or safety mitigations the developer will put in place when a model surpasses dangerous capability thresholds. The law explicitly identifies dangerous capabilities including CBRN weapons assistance, cyberattack capabilities, and scenarios where models might evade developer control, all of which directly parallel the risks identified in TC260's Framework. It also mandates 24 hour reporting for incidents posing "imminent risk of death or serious physical injury," enabling the state government to mount a rapid emergency response. SB53 took effect on January 1, 2026.

**New York Responsible AI Safety and Education (RAISE) Act**. The RAISE Act, signed into law in December 2025, would likewise require large AI developers to publish and implement frontier AI frameworks. These must "describe in detail" how the developer will evaluate whether their models have capabilities that could pose a "catastrophic risk" from misuse or loss of control, and to outline safeguards and procedures intended to mitigate such risks. It also mandates rapid 72-hour reporting of safety incidents so that state authorities can respond promptly. The RAISE Act has nearly identical text to SB53 though is strengthened in some areas; for example, repeat civil penalties extend up to three million dollars, compared to SB53's maximum of one million dollars. The law takes effect one year later than SB53, on January 1, 2027 [54].

### 3.3 Industry Best Practices and International Commitments

Beyond binding regulation, leading AI companies worldwide have also voluntarily created frameworks for managing catastrophic AI risks.

**AI Seoul Summit Commitments** (May 2024). Twenty companies—including Chinese companies like MiniMax, Zhipu.ai, and 01.ai alongside Western firms such as OpenAI, Anthropic, Google, Microsoft, and Meta—signed on to eight safety commitments [73]. The signatories commit to "assess the risks posed by their frontier models or systems across the AI lifecycle, including before deploying that model or system, and, as appropriate, before and during training." They must set "thresholds at which severe risks posed by a model or system, unless adequately mitigated, would be deemed intolerable" and establish "explicit processes they intend to follow if their model or system poses risks that meet or exceed the pre-defined thresholds." In extreme cases, signatories commit "not to develop or deploy a model or system at all, if mitigations cannot be applied to keep risks below the thresholds"—a strong pre-commitment that ensures emergency prevention takes precedence over commercial pressures to deploy. However, despite committing to publishing safety frameworks, none of the three Chinese company signatories have done so [44, 74]. Nor is this problem specific to Chinese companies: eight of the twenty total signatories have likewise failed to honor their commitment [44].

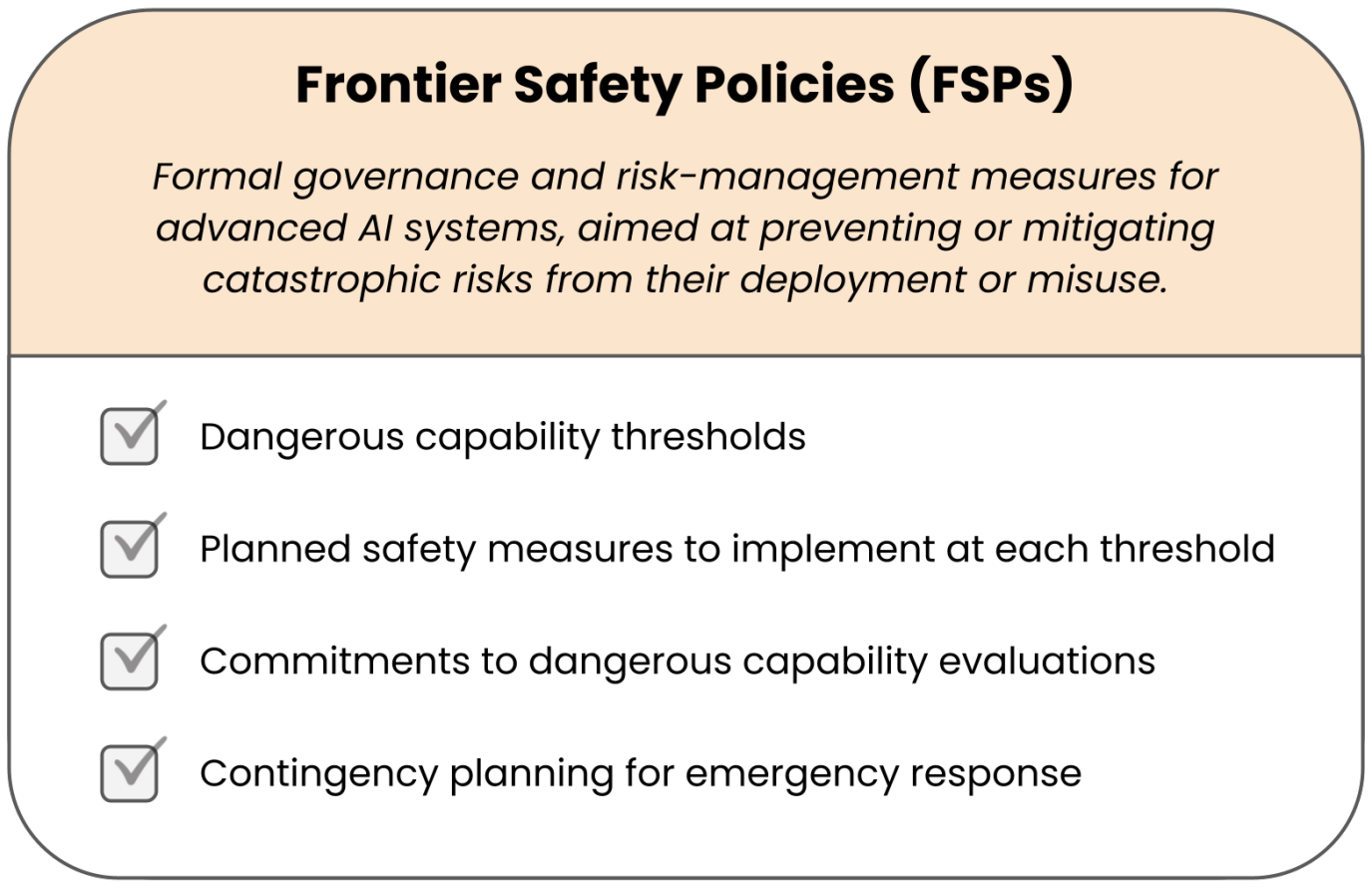

Fig. 1. Definition and key elements of a Frontier Safety Policy.

**Frontier Safety Policies (FSPs)**. AI companies have operationalized these commitments by writing detailed FSPs that create multiple layers of emergency preparedness [30, 43]. The definition and components of an FSP are shown in Figure 1. The core of an FSP is a taxonomy of dangerous capability thresholds, specific levels of AI capability that would pose severe risks if left unmitigated. Anthropic's Responsible Scaling Policy, for instance, defines thresholds for CBRN weapons engineering and autonomous AI research and development capabilities that would trigger increased security[2] [4]. Similarly, OpenAI's Preparedness Framework sets thresholds for biological and chemical capabilities, cybersecurity capabilities, and AI self-improvement capabilities [56]. These thresholds directly parallel the catastrophic risks identified in TC260's AI Safety Governance Framework (biological and chemical weapons, cyberattacks, and loss of control), suggesting an emerging global consensus.

FSPs often mandate continuous assessments to monitor whether models are approaching thresholds, in line with China's calls for early risk warning. When evaluations reveal that a model is approaching dangerous capability thresholds, FSPs prescribe escalating safeguards that mirror the "tiered management" approach advocated by China's State Council [14]. For models that pose only moderate risks, companies might implement enhanced input/output filtering or restrict API access. As capabilities approach more dangerous thresholds, the policies mandate increasingly stringent protections, such as pausing deployment of models too risky for public use.

In many cases, FSPs also mandate external transparency, especially transparency to government. Anthropic commits to "notify a relevant U.S. Government entity" of models requiring greater than ASL-2 security, and Google DeepMind will "share information with appropriate government authorities" when one of its models reaches a predefined critical capability level [33]. Such notification mechanisms contribute to early warning by ensuring that regulators know in advance when AI systems are approaching dangerous capability milestones, giving them time to prepare for possible incidents.

## 4 Implementations for the Four Phases of Emergency Response

Earlier sections have described how China's emerging AI governance framework already exhibits a foundation for emergency response in light of parallel developments in other jurisdictions. This section builds on that diagnosis by outlining a non-exhaustive set of possible implementation measures for AI emergency response across

[2]In May 2025, Anthropic activated its ASL-3 protections as a precautionary measure when launching Claude Opus 4, enhancing model weight security and making Claude Opus 4 more robust against jailbreaks, after determining that it could not clearly rule out that the model had reached the relevant capability threshold [3].

China's four-phase structure, while situating them among current practices. For a summary of the proposed implementations for each phase, see Table 1.

Table 1. China's four phases of emergency response with potential implementations.

| Emergency Response Phase | Goal | Implementations |
|---|---|---|
| Prevention and Preparedness | Reduce the risk of emergencies, and prepare response protocols | • Establish a tiered system of security and safety mitigations keyed to dangerous capability thresholds<br>• Share information about threats, vulnerabilities, and capabilities improvements among frontier AI companies<br>• Set up dedicated incident disclosure and communication channels between frontier model developers and relevant government authorities<br>• Develop standardized benchmarks to measure dangerous capabilities of frontier models |
| Monitoring and Early Warning | Detect imminent emergencies as early as possible | • Regularly evaluate models for dangerous capabilities, including through human uplift studies that measure marginal risk given availability of other tools<br>• Monitor continuously for dangerous or suspicious user queries<br>• Coordinate relevant authorities and compute providers to monitor for signs of rogue AI replication |
| Response and Rescue | Contain the emergency, and limit harm | • Rapidly remediate safety and security issues to the extent possible<br>• Suspend or restrict public and internal access to critically dangerous AI models<br>• Report incidents and risks to regulatory authorities and/or the public |
| Recovery and Reconstruction | Learn from successes and mistakes, and improve response protocols | • Conduct blameless postmortem investigations to identify failures<br>• Update safety policies and security mitigations in light of lessons learned<br>• Run regular tabletop exercises simulating past incidents and near-misses |

The **prevention and preparedness** phase aims to reduce the risk of emergencies before they occur. The primary implementation is for AI companies to adopt FSPs, tiered lists of dangerous capability thresholds with corresponding safety and security measures that the company commits to put in place if the threshold is met.

A second implementation is for AI companies to share information about emerging threats, model vulnerabilities, and expected capabilities improvements among themselves. This can enhance companies' awareness of risks and enable more timely response to impending emergencies. The members of the Frontier Model Forum—OpenAI, Google, Anthropic, Meta, Microsoft, and Amazon—have agreed to share risk and safety information, and other AI companies could join or form similar arrangements [29].

Third, companies and governments can establish dedicated incident disclosure and communication channels so that information can flow quickly in an emergency. Article 73 of the EU AI Act establishes a rapid incident reporting mechanism for providers of high-risk AI systems placed on the European market [3] [55]. California's Senate Bill 53 creates a similar mechanism for large AI developers to inform the state government of severe incidents involving their models [10].

A final prevention and preparedness implementation is for AI companies, government bodies, and civil society organizations to develop and share more standardized benchmarks to measure dangerous capabilities of frontier models. These benchmarks can strengthen the evidentiary basis for both internal safety evaluations and external oversight. They give all parties involved in governing frontier models a trusted and standardized metric by which to measure those models' dangerous capabilities.

The **monitoring and early warning** phase aims to detect unfolding emergencies as early as possible. The primary implementation discussed above is for AI companies, government regulators, and independent evaluators to regularly test models for dangerous capabilities. These evaluations could involve question-and-answer tests of dangerous knowledge, multistep agentic tasks, and human uplift studies. Frontier developers already use all three types of evaluations to assess whether their models are safe to deploy, as witnessed by recent model cards from Anthropic and OpenAI [5, 57].

AI companies that deploy their models via API or web app can detect emergencies early by monitoring their services continuously for suspicious queries that might indicate misuse. This mitigation has a track record of success. With continuous monitoring, OpenAI has detected and stopped automated scamming, influence operations, and malware campaigns that attempted to misuse their models [58]. Anthropic's Clio system has demonstrated that continuous query monitoring can be implemented while preserving user privacy [72]. Similarly, AI compute providers, with help from relevant authorities, can continuously monitor their clusters for signs of rogue AI replication during a loss of control event [69].

The **response and rescue** phase aims to limit harm by intervening rapidly once an emergency has begun. An AI company can respond to an unfolding emergency by revoking or restricting access to the models that are causing the emergency. This has been done for comparatively minor incidents in the past. For instance, Google temporarily paused Gemini's image generation of people in February 2024 after a content generation issue was discovered [61]. Compute providers can respond to loss of control emergencies by suspending or restricting access to the hardware resources models need to propagate autonomously [69]. Seeking help elsewhere, a company experiencing an emergency might be able to call upon pre-vetted external experts, from government or independent organizations, to help rapidly patch vulnerable models or advise the company in their response.

Finally, the **recovery and reconstruction** phase aims to improve response protocols for future emergencies by learning from successes and mistakes. One implementation widely used in industry is the blameless postmortem, an in-depth investigation that seeks to identify failures without seeking to punish or blame those involved [7]. Blamelessness helps to surface more valuable lessons by encouraging honest disclosures from people closest to the incident. AI companies should update their safety policies in light of what they learn from these postmortem investigations. AI companies, compute providers, and relevant government agencies can also keep the memory of past emergencies fresh by running regular tabletop simulations, so that staff remember emergency response protocols and lessons learned from previous close-calls.

[3] According to the Act's high-risk AI system framework on deployment context (Chapter III), as opposed to its GPAI systemic risk framework on model size and capability (Chapter V).

## 5 Related Work

This paper draws on two bodies of scholarship that have largely developed in isolation: research on frontier AI safety frameworks as a governance mechanism, and analysis of China's AI safety governance landscape. We review each in turn and then position our contribution. The primary regulatory instruments and industry initiatives themselves are described in detail in §2 and §3.

### 5.1 Scholarship on Frontier Safety Policies

In September 2023, Anthropic published the first frontier safety policy (named "Responsible Scaling Policy") [6], developed in collaboration with the Model Evaluation and Threat Research organization (METR). Later in the same month, METR published a broader proposal for responsible scaling policies [43]. Karnofsky [38] formalized the general structure as "if-then commitments": commitments of the form "if an AI model has capability X, risk mitigations Y must be in place." He argued that such commitments allow preparation for catastrophic AI risks without requiring agreement on whether those risks are imminent, and that voluntary adoption could prototype frameworks for future regulation. Karnofsky subsequently proposed specific "tripwire capabilities" that could serve as thresholds for enhanced safety measures [37].

The broader scholarly literature has examined frontier AI governance from several angles. Anderljung et al. [1] proposed a regulatory architecture for frontier AI centered on standard-setting, registration, and compliance mechanisms. Koessler and Schuett [39] reviewed risk assessment techniques from safety-critical industries including finance, aviation, nuclear power, and biolabs, assessing their applicability to frontier AI developers. Campos et al. [11] built on this work, drawing on practices from aviation, nuclear power, and other high-stakes sectors to propose a structured risk management process tailored to frontier AI.

Governance scholarship has also examined the institutional infrastructure needed to make these frameworks credible. Schuett [63] argued that frontier AI developers need internal audit functions to verify compliance with their own safety commitments, noting that the high cost of pausing development creates strong incentives for superficial compliance. Homewood et al. [35] proposed third-party compliance reviews as an external accountability mechanism. Buhl et al. [8] applied the concept of safety cases – already common in industries such as aviation and nuclear power – to frontier AI: structured arguments, supported by evidence, that a system is safe enough to deploy in a given context. A comparative analysis by the Oxford AI Governance Institute found that while published safety frameworks increasingly share a high-level structure, they often differ in terminology and specificity, and generally do not reference the broader risk management literature [85].

Somani et al. [69] applied emergency preparedness concepts directly to AI risks, developing detection, escalation, and containment protocols for AI loss-of-control scenarios, though focused on Western institutional contexts rather than China's emergency management architecture. As of December 2025, twelve companies had published frontier safety policies [44]. These frameworks have informed binding regulation in the EU, California, and New York (§3). However, the durability of voluntary commitments remains uncertain: Anthropic's February 2026 revision of its RSP dropped its commitment to pause model development if safety measures were inadequate, citing collective action problems [2]. METR's policy director called the change "more evidence that society is not prepared for the potential catastrophic risks posed by AI," expressing concern that abandoning "binary thresholds" could enable a "frog-boiling effect" of gradual escalation of risk over time [59]. This trajectory underscores the case for embedding FSP-like requirements in binding regulatory frameworks, as we propose for China's emergency response system.

### 5.2 Scholarship on Chinese AI Safety Governance

China's AI governance has attracted growing scholarly attention, though the literature has focused on different concerns than those motivating frontier safety policies. Zhu et al. [84], writing in *Science* with co-authors from DeepSeek and Alibaba, describe China's regulatory framework as structured around six pillars emphasizing open-source exemptions, efficient adjudication through Internet Courts, and experimentalist phasing-in of new rules; they argue that the framework has facilitated an "innovation- and openness-friendly institutional environment" but call for a consolidated AI law to provide greater clarity and harmonization. Zhang [83] analyzes the institutional dynamics behind China's rapid adoption of AI regulations, arguing in the *Columbia Journal of Transnational Law* that the Chinese government's simultaneous roles as policymaker, investor, supplier, customer, and regulator create tensions that shape the character of its AI rules. Zou et al. [86], writing in the *Cambridge Forum on AI: Law and Governance*, observe that Chinese AI regulation has received "comparatively little notice" internationally despite China being the first country to implement binding AI rules, and identify a common thread across these regulations: their emphasis on content safety rather than other risk categories. The Asia Society [16] notes that while no Chinese regulation has formally established a risk-based classification framework comparable to the EU AI Act, the government's implicit risk hierarchy places control of information as the primary concern.

Within China's academic system, the dominant governance framework is "agile governance" (敏捷治理). Xue, Jia, and Zhao [81] – Xue Lan chairs China's national AI governance expert committee – propose in *Chinese Public Administration* a classification system matching AI risk types to regulatory tools, arguing that bureaucratic regulators face inherent limitations in responding to rapid technological change. This paper, with over 80 citations, articulates the intellectual foundation for China's "classified and graded regulation" (分类分级监管) approach. In a related article in *Administrative Reform*, Xue and Wang [79] identify five governance challenges intensified by frontier AI development, including the gap between technological development and governance capacity. Zeng, Liang, and Zhang [82] trace the evolution from "soft law first" to soft-hard law coordination in Chinese AI governance, a dynamic relevant to understanding how non-binding TC260 frameworks may eventually become mandatory standards.

Catastrophic risk from frontier AI has entered Chinese policy discourse recently through non-binding documents rather than binding regulation. The Carnegie Endowment's analysis of TC260's AI Safety Governance Framework 2.0 characterizes the September 2025 update as a specific official recognition that AI systems inevitably absorb dual-use knowledge during training, quoting the Framework's warning about "loss of control over knowledge and capabilities of nuclear, biological, chemical, and missile weapons" [66]. Concordia AI's survey of Chinese AI developers found that only three of thirteen published risk evaluation results, documenting a persistent gap between voluntary commitments and practice [74]. Calero [9] traces how Chinese governance proposals have evolved from ambitious but vague early plans to more specific recent instruments that assign defined roles and responsibilities to governance actors, though ambiguity remains even in the latest proposals.

The rapid adoption of the open-source AI agent OpenClaw in early 2026 has provided an early, concrete signal of several risks identified in this literature. After OpenClaw gained widespread use across Chinese tech companies and local governments, central regulators moved swiftly to restrict its deployment at state-owned enterprises and government agencies, citing data security concerns [62]. CNCERT warned that "due to a misunderstanding of user operation instructions and intentions, OpenClaw may completely delete important information such as emails and core production data" [45], a scenario that represents a limited but real instance of autonomous AI action deviating from user intent. The Ministry of Industry and Information Technology's

National Vulnerability Database (NVDB) further cautioned that OpenClaw's "blurred trust boundaries" during deployment, combined with its "continuous operation, autonomous decision-making, and the ability to invoke system and external resources," could lead to "unauthorized operations due to misleading instructions, configuration defects, or malicious takeover" [53]. In response, the AIIA issued detailed security guidance calling on organizations to "establish a real-time monitoring and emergency response mechanism for abnormal behavior, and shut down, handle, trace, and rectify any risks as soon as they are discovered" [19]. While these unintended actions attributed to OpenClaw fall short of the catastrophic loss-of-control outcomes discussed in this paper, the episode illustrates how agentic AI systems with broad permissions can produce harmful autonomous behavior, and how China's existing regulatory apparatus has responded with measures that parallel some of the emergency response mechanisms analyzed here.

Comparative scholarship has examined convergences between Chinese and Western approaches to AI risk. Siddiqui et al. [68], in a systematic analysis of over 40 policy documents from both nations in their original languages, find that U.S. labs define and test for "dangerous capabilities" in their safety frameworks, whereas Chinese labs acknowledge these practices but have not yet systematically adopted them. Wasil and Durgin [76] compile public statements from leaders in both countries on extreme AI risks, finding concern about AGI, intelligence explosions, and loss of control expressed in each. The AI Frontiers project observes that the Shanghai AI Laboratory's frontier risk evaluations show "closer convergence with risk evaluations conducted in the U.S. and UK," employing shared evaluation methods, suggesting that certain technical safety practices may be converging even where regulatory frameworks have not [26].

### 5.3 Positioning This Paper

Prior work on frontier safety policies has focused predominantly on Western companies and regulatory contexts. The scholarship on Chinese AI governance has examined domestic instruments in depth but has given limited attention to how the FSP model could be adapted for China. Our paper bridges these literatures by analyzing how FSPs – developed through Western industry practice and now codified in Western regulation – map onto China's established emergency response architecture. To our knowledge, this is the first paper to systematically analyze the integration of frontier safety policies into China's four-phase emergency management framework, examining how existing generative AI requirements could be extended to mandate dangerous capability evaluations and tiered safety mitigations.

## 6 Key Considerations and Future Directions

Top Chinese authorities have recognized the need for AI emergency preparedness, and regulators have laid a strong foundation for future rules to manage AI emergencies. To strengthen preparedness for catastrophic AI risks, the generative AI registry approval could be conditioned on companies both adopting FSPs and passing dangerous capability evaluations as part of pre-deployment assessment.

**The Case for Frontier Safety Policies (FSPs)**. The strategic goal of AI emergency preparedness requires implementing all four phases of China's established emergency response framework: prevention and preparedness, monitoring and early warning, response and rescue, and recovery and reconstruction. Frontier safety policies provide concrete technical mechanisms for the first two phases, which must be established *before* any emergency occurs. For prevention and preparedness, committing to FSPs would require AI developers to define dangerous capability thresholds and pre-plan safety mitigations, ensuring they are ready with specific countermeasures before dangerous capabilities emerge. This aligns with the Chinese governance principle that effective emergency management begins with proactive prevention rather than post-incident response [40].

For monitoring and early warning, committing to FSPs would incorporate more continuous capability evaluations that serve as early indicators of emerging risks. Just as the China Earthquake Administration monitors seismic activity to provide early earthquake warnings [21], AI developers must monitor their models' capabilities to detect when they approach dangerous thresholds. This way, monitoring does not wait until after an AI system demonstrates harmful capabilities in the real world, by which point an emergency may already be underway. These proactive phases enable effective response and rescue when needed. Pre-defined thresholds trigger pre-planned mitigations automatically, avoiding the chaos of formulating responses during a crisis. Additionally, the systematic evaluation data collected through FSP commitments would support post-incident recovery and reconstruction by providing clear records of what capabilities emerged when and how the response unfolded.

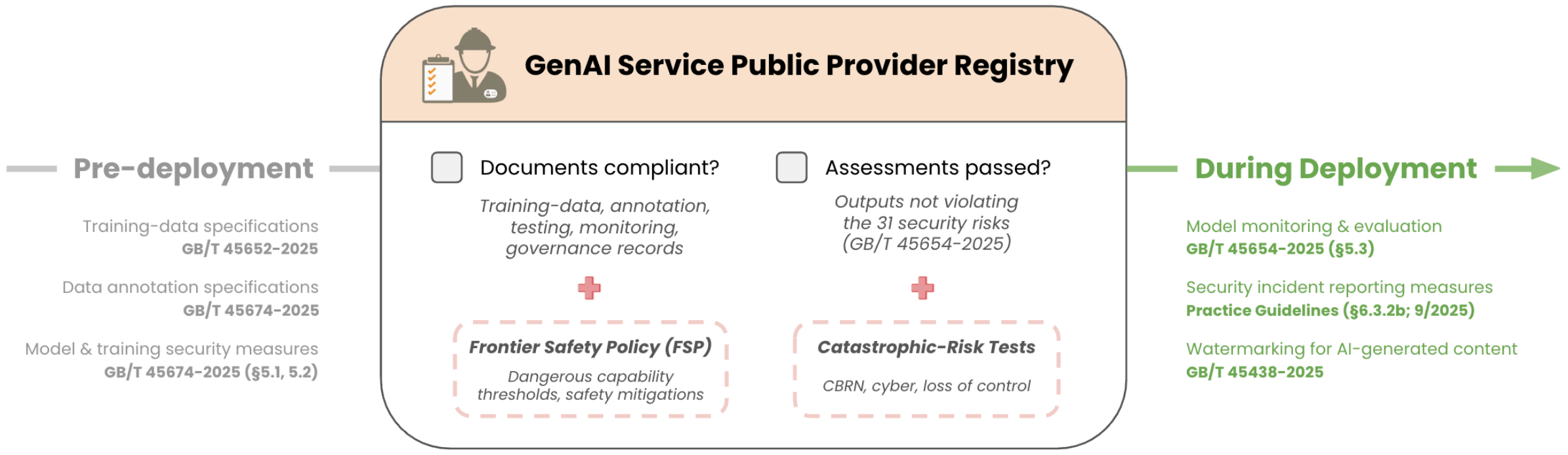

Fig. 2. Registry-centric model lifecycle illustrating China's current regulatory ecosystem alongside our proposed additions: requiring providers to develop Frontier Safety Policies (FSPs) and integrating a catastrophic-risk test suite into pre-deployment assessments.

**Testing for Catastrophic AI Risks**. FSPs are a suitable candidate for implementing key emergency planning goals. Some elements of FSPs are already present in TC260's Basic Security Requirements (GB/T 45654-2025), which direct generative AI model developers to evaluate their models' outputs against content security requirements before deployment and to establish plans for managing security emergencies [47]. Expanding the scope of required pre-deployment testing to include dangerous capability evaluations would extend these requirements to address catastrophic risks. For example, model developers might be required to run the Shanghai AI Laboratory's suite of frontier risk evaluations on all of their new frontier models and report evaluation results [65]. Future regulation might leave it up to AI developers to set their own dangerous capability thresholds in the spirit of SB53, or they might require specific mitigations at designated risk thresholds.

**Envisioning Extensions to Existing Generative AI Requirements**. Figure 2 shows how existing requirements for generative AI service providers could be modified to mandate FSPs. Under GB/T 45654-2025, AI service providers must pass two steps before offering a model to the public [47]. First, they must maintain documents demonstrating that the model was trained on a compliant data set, that the company followed data annotation rules, and that they have a sound plan for monitoring during deployment. Second, the company's model must pass assessments to verify that it does not pose any of the 31 security risks. The documentation requirements could be augmented by mandating providers maintain a frontier safety policy alongside other required records. At a minimum, this FSP should identify several dangerous capability thresholds beyond the current model's capability level and specify what additional safety mitigations the provider will implement when those thresholds are met. The security assessment requirements could be augmented by adding evaluations for catastrophic risk to the existing suite of assessments.

Indeed, there is precedent for these changes in preliminary drafts of the Basic Security Requirements. A draft released in February 2024 directed model providers to "pay close attention to the long-term risks that may be brought about by generative AI," and specifically mentioned risks of AI self-replicating, writing malware, and making biological or chemical weapons [49]. TC260's prior work reflects early consideration of catastrophic AI risks and initial threat modeling efforts. The Committee could build on this foundation by developing a new standard that operationalizes its earlier goals through implementing FSPs or similar mechanisms.

Chinese authorities could also write an obligation for frontier AI developers to adopt FSPs into future legislation. China's national legislature has yet to pass laws intended to regulate AI, but the State Council has repeatedly indicated that such laws are being drafted [31, 32]. One or more of the laws that eventually emerge could contain language requiring frontier AI companies to evaluate their models for catastrophic risks before deployment and to pre-plan tiered safety measures. This would be consistent with existing scholarly proposals for China's AI law, such as the Model Artificial Intelligence Law put forward by the Chinese Academy of Social Sciences [22]. Their proposal would require AI providers to assess risks stemming from their models before and during deployment and to manage those risks through pre-established "safety risk management systems," capturing some of the key aspects of FSPs. A law like this might set China's AI emergency response regime on a sturdier legal foundation than regulatory standards alone can provide.

China's regulatory architecture and industry frameworks provide a foundation for more comprehensive AI emergency preparedness. Chinese AI companies have begun committing to voluntary safety agreements, and existing requirements already feature FSP-like qualities. By mandating frontier safety policies and dangerous capability evaluations, China would formalize these emerging practices to address catastrophic AI risks. Such measures would represent a significant step in the country's evolving approach to frontier AI risk governance.

**Limitations**. While FSPs offer a promising approach to emergency preparedness, they have inherent limitations. They primarily address risks that can be anticipated and measured through capability evaluations, and may be less effective against the most novel and unforeseeable threats. Their effectiveness depends critically on the quality of evaluations, the appropriateness of thresholds, and companies' adherence to their commitments. Moreover, our proposed integration of FSPs into China's regulatory framework combines corporate self-auditing with government oversight, each carrying its own incentive-related risks: government-led systems may introduce rigid requirements, while self-governance can lead to opacity or underreporting. FSPs also focus primarily on the prevention and monitoring phases of emergency response. The response and recovery phases require additional implementations, as we discuss in Section 4. Despite these limitations, FSPs still offer a concrete starting point for operationalizing the proactive stages of an emergency response framework.

## Acknowledgments

James was supported by the ERA Fellowship, and Miles was supported by the Centre for the Governance of AI. We would like to thank both organizations for their financial and intellectual support. We would also like to thank Aris Richardson, Saad Siddiqui, Jason Zhou, Miro Pluckebaum, and Cameron Tice for helpful discussions.

## Generative AI Usage Statement

Generative AI (ChatGPT 4.5; Claude Sonnet 4, 4.6; Claude Opus 4, 4.6) was used to improve LaTeX table formatting, assist literature review, and translate source documents from Chinese to English.

## References

[1] Anderljung, Markus, Joslyn Barnhart, Anton Korinek, Jade Leung, Cullen O'Keefe, Jess Whittlestone, et al. (2023). Frontier AI Regulation: Managing Emerging Risks to Public Safety. https://arxiv.org/abs/2307.03718

[2] Anthropic. (February 2026). Responsible Scaling Policy Version 3.0. https://www.anthropic.com/news/responsible-scaling-policy-v3

[3] Anthropic. (May 2025). Activating AI Safety Level 3 protections. https://www.anthropic.com/news/activating-asl3-protections

[4] Anthropic. (May 2025). Responsible Scaling Policy Version 2.2. https://www-cdn.anthropic.com/872c653b2d0501d6ab44cf87f43e1dc4853e4d37.pdf

[5] Anthropic. (May 2025). System Card: Claude Opus 4 & Claude Sonnet 4. https://www-cdn.anthropic.com/07b2a3f9902ee19fe39a36ca638e5ae987bc64dd.pdf

[6] Anthropic. (September 2023). Responsible Scaling Policy, Version 1.0. https://www-cdn.anthropic.com/1adf000c8f675958c2ee23805d91aaade1cd4613/responsible-scaling-policy.pdf

[7] Beyer, Betsy, Chris Jones, Jennifer Petoff, & Niall Richard Murphy, eds. (2017). *Site Reliability Engineering.*

[8] Buhl, Marie Davidsen, Gaurav Sett, Leonie Koessler, Jonas Schuett, & Markus Anderljung. (2024). Safety cases for frontier AI. https://arxiv.org/abs/2410.21572

[9] Calero, Hipolito. (September 2024). An Analysis of China's AI Governance Proposals. Center for Security and Emerging Technology. https://cset.georgetown.edu/article/an-analysis-of-chinas-ai-governance-proposals/

[10] California State Legislature. (2025). Senate Bill 53. https://leginfo.legislature.ca.gov/faces/billTextClient.xhtml?bill_id=202520260SB53

[11] Campos, Simeon, Henry Papadatos, Fabien Roger, Chloé Touzet, Otter Quarks, & Malcolm Murray. (2025). A frontier AI risk management framework: Bridging the gap between current AI practices and established risk management. https://arxiv.org/abs/2502.06656

[12] Center for International Security and Strategy, Tsinghua University. (2025). Glossary Research on Artificial Intelligence Risks (Part I). https://ciss.tsinghua.edu.cn/info/eyjdt/8078

[13] Center for Security and Emerging Technology. (2024). Translation: Basic Safety Requirements for Generative AI Services. https://cset.georgetown.edu/publication/china-safety-requirements-for-generative-ai-final/

[14] Central Committee of the Communist Party of China & State Council. (2025). National Emergency Response Plan. https://www.gov.cn/zhengce/202502/content_7005635.htm

[15] Charlie Bullock (2025). Legal Issues Raised by the Proposed Executive Order on AI Preemption. https://law-ai.org/legal-issues-raised-by-the-proposed-executive-order-on-ai-preemption/

[16] Chen, Qiheng. (2024). China's Emerging Approach to Regulating General-Purpose Artificial Intelligence: Balancing Innovation and Control. https://asiasociety.org/policy-institute/chinas-emerging-approach-regulating-general-purpose-artificial-intelligence-balancing-innovation-and

[17] China Academy of Information and Communications Technology. (2025). Artificial Intelligence Safety Research Report: Security Risk Analysis and Response from a Technical Perspective. https://aihub.caict.ac.cn/f/d/48c0bbc85c3948a62b615ceac6e9c1ce

[18] China AI Industry Alliance (AIIA). (December 2024). Artificial Intelligence Safety Commitments. https://mp.weixin.qq.com/s/s-XFKQCWhu0uye4opgb3Ng

[19] China AI Industry Development Alliance (AIIA). (March 2026). 关于防范OpenClaw开源AI智能体安全风险的提示 (Notice on Preventing Security Risks of OpenClaw Open-Source AI Agent). https://mp.weixin.qq.com/s/gmaqDpVnf9jtJUN4qJpA6g

[20] China AI Safety and Development Association. (July 2025). China Artificial Intelligence Security and Safety Commitments Framework. https://mp.weixin.qq.com/s/jwtALdsfXLMN_ohfsOF5Dw

[21] China Earthquake Administration. (2024). China establishes world's largest quake early warning network. https://english.www.gov.cn/news/202407/26/content_WS66a3a393c6d0868f4e8e97cc.html

[22] Chinese Academy of Social Sciences Research Group on AI Ethics and Governance. (2023). Model Artificial Intelligence Law 1.1. http://iolaw.cssn.cn/zxzp/202309/W020230907361599893636.pdf

[23] Chow, Vincent (2025). DeepSeek evaluates AI models for 'frontier risks', source says, as China promotes safety. https://www.scmp.com/tech/article/3325742/deepseek-evaluates-ai-models-frontier-risks-source-says-china-promotes-safety

[24] Cyberspace Administration of China (CAC). (2023). Interim Measures for the Management of Generative Artificial Intelligence Services. https://www.cac.gov.cn/2023-07/13/c_1690898327029107.htm

[25] Ding, Zhang. AI Safety Governance Under the Global Security Initiative (人工智能安全治理). https://ciss.tsinghua.edu.cn/info/rgzn_yjcg/6806

[26] Elmgren, Karson, Scott Singer, & Oliver Guest. (October 2025). Is China Serious About AI Safety? AI Frontiers. https://ai-frontiers.org/articles/is-china-serious-about-ai-safety

[27] European Commission. (2025). Code of Practice for General-Purpose AI Models. https://digital-strategy.ec.europa.eu/en/policies/contents-code-gpai

[28] Executive Office of the President. (December 2025). Executive Order on Ensuring a National Policy Framework for Artificial Intelligence. https://www.whitehouse.gov/presidential-actions/2025/12/eliminating-state-law-obstruction-of-national-artificial-

intelligence-policy/
[29] Frontier Model Forum. (2025). FMF Announces First-of-its-Kind Information-Sharing Agreement. https://www.frontiermodelforum.org/updates/fmf-announces-first-of-its-kind-information-sharing-agreement/
[30] Frontier Model Forum. (2024). Issue Brief: Components of Frontier AI Safety Frameworks. https://www.frontiermodelforum.org/updates/issue-brief-components-of-frontier-ai-safety-frameworks/
[31] General Office of the State Council. (May 2023). Notice of the General Office of the State Council on Issuing the 2023 Legislative Work Plan of the State Council. https://www.gov.cn/zhengce/content/202306/content_6884925.htm
[32] General Office of the State Council. (May 2024). Notice of the General Office of the State Council on Issuing the 2024 Legislative Work Plan of the State Council. https://www.gov.cn/zhengce/content/202405/content_6950093.htm
[33] Google DeepMind. (February 2025). Frontier Safety Framework Version 2. https://storage.googleapis.com/deepmind-media/DeepMind.com/Blog/updating-the-frontier-safety-framework/Frontier%20Safety%20Framework%202.0.pdf
[34] He, Alex. (November 2025). AI Development and Governance in China. CIGI Paper No. 338. Centre for International Governance Innovation. https://www.cigionline.org/documents/3593/no.338_He.pdf
[35] Homewood, Aidan, Sophie Williams, Noemi Dreksler, John Lidiard, Malcolm Murray, Lennart Heim, Marta Ziosi, Seán Ó hÉigeartaigh, Michael Chen, Kevin Wei, Christoph Winter, Miles Brundage, Ben Garfinkel, & Jonas Schuett. (2025). Third-party compliance reviews for frontier AI safety frameworks. https://arxiv.org/abs/2505.01643
[36] International Dialogues on AI Safety. (2024). The global nature of AI risks makes it necessary to recognize AI safety as a global public good. https://idais.ai/dialogue/idais-venice/
[37] Karnofsky, Holden. (December 2024). A Sketch of Potential Tripwire Capabilities for AI. Carnegie Endowment for International Peace. https://carnegieendowment.org/research/2024/12/a-sketch-of-potential-tripwire-capabilities-for-ai
[38] Karnofsky, Holden. (September 2024). If–Then Commitments for AI Risk Reduction. Carnegie Endowment for International Peace. https://carnegieendowment.org/research/2024/09/if-then-commitments-for-ai-risk-reduction
[39] Koessler, Leonie & Jonas Schuett. (2023). Risk assessment at AGI companies: A review of popular risk assessment techniques from other safety-critical industries. https://arxiv.org/abs/2307.08823
[40] Lan, Xue, Shen Hua. (August 2021). Five Major Transformations: Updating Concepts for Emergency Management System Construction in the New Era. https://ccmr.sppm.tsinghua.edu.cn/papers/1160.jhtml
[41] Laurent, Jon M., Joseph D. Janizek, Michael Ruzo, Michaela M. Hinks, Michael J. Hammerling, Siddharth Narayanan, Manvitha Ponnapati, Andrew D. White, Samuel G. Rodriques. (July 2024). LAB-Bench: Measuring Capabilities of Language Models for Biology Research. https://arxiv.org/abs/2407.10362
[42] Li, Nathaniel, Alexander Pan, Anjali Gopal, Summer Yue, Daniel Berrios, Alice Gatti, Justin D. Li, Ann-Kathrin Dombrowski, Shashwat Goel, Long Phan, Gabriel Mukobi, Nathan Helm-Burger, Rassin Lababidi, Lennart Justen, Andrew B. Liu, Michael Chen, Isabelle Barrass, Oliver Zhang, Xiaoyuan Zhu, Rishub Tamirisa, Bhrugu Bharathi, Adam Khoja, Zhenqi Zhao, Ariel Herbert-Voss, Cort B. Breuer, Samuel Marks, Oam Patel, Andy Zou, Mantas Mazeika, Zifan Wang, Palash Oswal, Weiran Liu, Adam A. Hunt, Justin Tienken-Harder, Kevin Y. Shih, Kemper Talley, John Guan, Russell Kaplan, Ian Steneker, David Campbell, Brad Jokubaitis, Alex Levinson, Jean Wang, William Qian, Kallol Krishna Karmakar, Steven Basart, Stephen Fitz, Mindy Levine, Ponnurangam Kumaraguru, Uday Tupakula, Vijay Varadharajan, Yan Shoshitaishvili, Jimmy Ba, Kevin M. Esvelt, Alexandr Wang, Dan Hendrycks. (May 2024). The WMDP Benchmark: Measuring and Reducing Malicious Use With Unlearning. https://www.wmdp.ai/
[43] Model Evaluation and Threat Research. (2023). Responsible Scaling Policies. https://metr.org/blog/2023-09-26-rsp/
[44] Model Evaluation and Threat Research. (December 2025). Common Elements of Frontier AI Safety Policies. https://metr.org/blog/2025-12-09-common-elements-of-frontier-ai-safety-policies/
[45] National Computer Network Emergency Response Technical Team/Coordination Center of China (CNCERT). (March 2026). 关于OpenClaw安全应用的风险提示 (Risk Alert on the Safe Use of OpenClaw). https://www.cert.org.cn/publish/main/11/2026/20260312144519429724511/20260312144519429724511_.html
[46] National Information Security Standardization Technical Committee (TC260) & National Computer Network Emergency Response Technical Team (September 2025). AI Safety Governance Framework 2.0. https://www.cac.gov.cn/2025-09/15/c_1759653448369123.htm
[47] National Information Security Standardization Technical Committee (TC260). (April 2025). Basic Security Requirements for Generative Artificial Intelligence Services (GB/T 45654-2025). https://www.tc260.org.cn/upload/2025-06-30/1751257342816036759.pdf
[48] National Information Security Standardization Technical Committee (TC260). (December 2024). Generative Artificial Intelligence Service Security Emergency Response Guidelines (Draft). https://www.tc260.org.cn/upload/2024-12-18/1734483139154029117.pdf
[49] National Information Security Standardization Technical Committee (TC260). (February 2024). Basic Security Requirements for Generative Artificial Intelligence Services (Draft). https://cset.georgetown.edu/publication/china-safety-requirements-for-generative-ai-final/
[50] National Information Security Standardization Technical Committee (TC260). (January 2025). AI Safety Standards System V1.0 (Draft). https://www.tc260.org.cn/upload/2025-01-24/1737709785951070331.pdf

[51] National Information Security Standardization Technical Committee (TC260). (September 2024). AI Safety Governance Framework. https://www.tc260.org.cn/upload/2024-09-09/1725849192841090989.pdf

[52] National Information Security Standardization Technical Committee (TC260). (September 2025). Cybersecurity Standard Practice Guidelines – Security Emergency Response Guidelines for Generative Artificial Intelligence Services https://www.tc260.org.cn/upload/2025-09-15/1757913511246034883.pdf

[53] National Vulnerability Database (NVDB), Ministry of Industry and Information Technology. (February 2026). 关于防范OpenClaw开源AI智能体安全风险的预警提示 (Early Warning on Preventing Security Risks of OpenClaw Open-Source AI Agent). https://mp.weixin.qq.com/s/2JWBD1tCPn4RStSZaVnbiw

[54] New York State Legislature. (2026). Assembly Bill A09449, Responsible AI Safety and Education Act. https://nyassembly.gov/leg/?default_fld=&leg_video=&bn=A09449&term=2025&Summary=Y&Text=Y

[55] Official Journal. (2024). Regulation (EU) 2024/1689 of the European Parliament and of the Council.

[56] OpenAI. (April 2025). Preparedness Framework Version 2. https://cdn.openai.com/pdf/18a02b5d-6b67-4cec-ab64-68cdfbddebcd/preparedness-framework-v2.pdf

[57] OpenAI. (August 2025). GPT-5 System Card. https://openai.com/index/gpt-5-system-card/

[58] OpenAI. (June 2025). Disrupting Malicious uses of AI: June 2025. https://cdn.openai.com/threat-intelligence-reports/5f73af09-a3a3-4a55-992e-069237681620/disrupting-malicious-uses-of-ai-june-2025.pdf

[59] Perrigo, Billy. (February 2026). Exclusive: Anthropic Drops Flagship Safety Pledge. *TIME*. https://time.com/7380854/exclusive-anthropic-drops-flagship-safety-pledge/

[60] Phuong, Mary, Matthew Aitchison, Elliot Catt, et al. (2024). Evaluating Frontier Models for Dangerous Capabilities. https://arxiv.org/abs/2403.13793

[61] Raghavan, Prabhakar. (2024). Gemini image generation got it wrong. We'll do better. https://blog.google/products/gemini/gemini-image-generation-issue/

[62] Reuters. (March 2026). China warns state-owned firms and government agencies against OpenClaw AI, sources say. https://www.reuters.com/technology/china-moves-curb-use-openclaw-ai-banks-state-agencies-bloomberg-news-reports-2026-03-11/

[63] Schuett, Jonas. (2025). Frontier AI developers need an internal audit function. *Risk Analysis*. https://doi.org/10.1111/risa.17665

[64] Shanghai Artificial Intelligence Laboratory & Concordia AI. (2025). Frontier AI Risk Management Framework (v1.0). https://concordia-ai.com/research/frontier-ai-risk-management-framework/

[65] Shanghai Artificial Intelligence Laboratory. (July 2025). Frontier AI Risk Management Framework in Practice: A Risk Analysis Technical Report. https://arxiv.org/abs/2507.16534

[66] Sheehan, Matt & Scott Singer. (October 2025). How China Views AI Risks and What to do About Them. https://carnegieendowment.org/research/2025/10/how-china-views-ai-risks-and-what-to-do-about-them

[67] Shevlane, Toby, Sebastian Farquhar, Ben Garfinkel, Mary Phuong, Jess Whittlestone, Jade Leung, et al. (2023). Model evaluation for extreme risks. https://arxiv.org/abs/2305.15324

[68] Siddiqui, Saad, Lujain Ibrahim, Kristy Loke, Stephen Clare, Marianne Lu, Aris Richardson, Conor McGlynn, & Jeffrey Ding. (2025). Promising Topics for U.S.–China Dialogues on AI Risks and Governance. In *Proceedings of ACM FAccT 2025*. https://arxiv.org/abs/2505.07468

[69] Somani, Elika, Anjay Freedman, Henry Wu, Marianne Lu, Chris Byrd, Henri van Soest, & Sana Zakaria. (2025). Strengthening Emergency Preparedness and Response for AI Loss of Control Events. https://www.rand.org/pubs/research_reports/RRA3847-1.html

[70] State Council of the People's Republic of China. (2024). National Natural Disaster Relief Emergency Plan. https://www.gov.cn/zhengce/content/202402/content_6930038.htm

[71] State Council of the People's Republic of China. (2025). China's National Security in the New Era. http://politics.people.com.cn/n1/2025/0512/c1001-40478167.html

[72] Tamkin, Alex, Miles McCain, Kunal Handa, Esin Durmus, Liane Lovitt, Ankur Rathi, Saffron Huang, Alfred Mountfield, Jerry Hong, Stuart Ritchie, Michael Stern, Brian Clarke, Landon Goldberg, Theodore R. Sumers, Jared Mueller, William McEachen, Wes Mitchell, Shan Carter, Jack Clark, Jared Kaplan, & Deep Ganguli. (December 2024). Clio: Privacy-Preserving Insights into Real-World AI Use. https://arxiv.org/abs/2412.13678v1

[73] United Kingdom Department for Science, Innovation and Technology. (2024). Frontier AI Safety Commitments. https://www.gov.uk/government/publications/frontier-ai-safety-commitments-ai-seoul-summit-2024/frontier-ai-safety-commitments-ai-seoul-summit-2024

[74] Wagner, Gabriel, Jason Zhou, Kwan Yee Ng, & Brian Tse. (July 2025). State of AI Safety in China (2025). Concordia AI. https://concordia-ai.com/research/state-of-ai-safety-in-china-2025/

[75] Wang, Zhe, Emily Chan, Kevin Liu, & May Yeung. (2016). The Disaster and Emergency Management System in China. https://www.hkjcdpri.org.hk/download/policy/PolicyBriefDisasterandEmergMxSysinChina.pdf

[76] Wasil, Akash & Tim Durgin. (2024). US-China perspectives on extreme AI risks and global governance. https://arxiv.org/abs/2407.16903

[77] Xinhua. (2025). Decision of the Standing Committee of the National People's Congress on Amending the Cybersecurity Law of the People's Republic of China. https://www.gov.cn/yaowen/liebiao/202510/content_7046194.htm

[78] Xinhua. (2025). Xi urges promoting healthy and orderly development of AI. http://english.www.gov.cn/news/202504/29/content_WS68100ef1c6d0868f4e8f2275.html

[79] Xue, Lan & Jingyu Wang (薛澜、王净宇). (2024). Frontier Trends in AI Development, Governance Challenges and Response Strategies (人工智能发展的前沿趋势、治理挑战与应对策略). *Administrative Reform* (行政管理改革), 2024(8). https://ciss.tsinghua.edu.cn/upload_files/atta/1728607551496_C0.pdf

[80] Xue, Lan & Kai Jia. (2024). AI Ethics and Safety Risk Governance: A Global Comparison and Chinese Practice (人工智能伦理问题与安全风险治理的全球比较与中国实践). *Public Management Review* (公共管理评论). http://cistp.sppm.tsinghua.edu.cn/info/1028/1168.htm

[81] Xue, Lan, Kai Jia, & Jing Zhao (薛澜、贾开、赵静). (2024). Agile Governance of AI in Practice: Classified Regulation and Policy Toolkit (人工智能敏捷治理实践:分类监管思路与政策工具箱构建). *Chinese Public Administration* (中国行政管理), 2024(3):99–110. https://hub.baai.ac.cn/view/36560

[82] Zeng, Xiong, Zheng Liang, & Hui Zhang (曾雄、梁正、张辉). (2024). Optimizing AI Soft Law Governance: From Soft Law First to Soft and Hard Law Coordination (人工智能软法治理的优化进路:由软法先行到软法与硬法协同). *E-Government* (电子政务), 2024(6):96–107. https://aiig.tsinghua.edu.cn/info/1368/2071.htm

[83] Zhang, Angela Huyue. (2025). The Promise and Perils of China's Regulation of Artificial Intelligence. *Columbia Journal of Transnational Law*. https://www.jtl.columbia.edu/volume-63/the-promise-and-perils-of-chinas-regulation-of-artificial-intelligence

[84] Zhu, Yue, Bo He, Hongyu Fu, Naying Hu, Shaoqing Wu, Taolue Zhang, Xinyi Liu, Gang Xu, Linghan Zhang, & Hui Zhou. (2025). China's emerging regulation toward an open future for AI. *Science*, 390(6769), 132–135. https://doi.org/10.1126/science.ady7922

[85] Ziosi, Marta, James Gealy, Miro Pluckebaum, Daniel Kossack, Simeon Campos, Lama Saouma, Uzma Chaudhry, Lisa Soder, Merlin Stein. (October 2025). Safety Frameworks and Standards: A Comparative Analysis to Advance Risk Management of Frontier AI. Research Memo. https://aigi.ox.ac.uk/wp-content/uploads/2025/10/Post-convening-memo_-Safety-Frameworks-and-Standards_-A-comparative-analysis-to-advance-risk-management-of-frontier-AI_09.10.2025.pdf

[86] Zou, Mimi et al. (2025). Navigating China's regulatory approach to generative artificial intelligence and large language models. *Cambridge Forum on AI: Law and Governance*, 1, e8, 1–16. https://www.cambridge.org/core/journals/cambridge-forum-on-ai-law-and-governance/article/navigating-chinas-regulatory-approach-to-generative-artificial-intelligence-and-large-language-models/969B2055997BF42DE693B7A1A1B4E8BA